\documentclass[sigconf]{acmart}

\AtBeginDocument{%
	}

\setcopyright{acmcopyright}
\copyrightyear{2023}
\acmYear{2023}
\acmDOI{XXXXXXX.XXXXXXX}

\acmConference[Conference WSDM '16]{The Sixteen ACM International Conference on Web Search and Data Mining}{Feb 27-- Mar 2, 2023}{Singapore}

\begin{document}

\title{Multi-Feature Integration for Perception-Dependent Examination-Bias Estimation}


\author{
	Xiaoshu Chen, Xiangsheng Li, Kunliang Wei, Bin Hu, Lei Jiang, Zeqian Huang and Zhanhui Kang}
\affiliation{
	Tencent Machine Learning Platform Search, \\
	Shenzhen, Guangdong, China
}
\email{xschenranker6@gmail.com}

\def\authors{Xiaoshu Chen, Xiangsheng Li, Kunliang Wei, Bin Hu, Lei Jiang, Zeqian Huang and Zhanhui Kang}

\renewcommand{\shortauthors}{Chen et al.}

\begin{abstract}
Eliminating examination bias accurately is pivotal to apply click-through data to train an unbiased ranking model. However, most examination-bias estimators are limited to the hypothesis of Position-Based Model (PBM), which supposes that the calculation of examination bias only depends on the rank of the document. Recently, although some works introduce information such as clicks in the same query list and contextual information when calculating the examination bias, they still do not model the impact of document representation on search engine result pages (SERPs) that seriously affects one's perception of document relevance to a query when  examining. Therefore, we propose a Multi-Feature Integration Model (MFIM) where the examination bias depends on the representation of document except the rank of it. Furthermore, we mine a key factor slipoff counts that can indirectly reflects the influence of all perception-bias factors. Real world experiments on Baidu-ULTR dataset demonstrate the superior effectiveness and robustness of the new approach. The source code is available at \href{https://github.com/lixsh6/Tencent_wsdm_cup2023/tree/main/pytorch_unbias}{https://github.com/lixsh6/Tencent\_wsdm\_cup2023}
\end{abstract}




\maketitle

\section{Introduction}
Learning to rank is a crucial part of information retrieval system \cite{qin2010letor}. In practice, the ranking model is often trained by the user's implicit feedback, e.g. user clicks. However, there are usually many complex biases such as position bias \cite{joachims2017unbiased} in the click-through data. Therefore, Unbiased learning to rank (ULTR), dedicating to train a unbiased ranking model from such biased click-through data, has gained a lot of attention.

Currently, most of ultr models \cite{10.1145/3534678.3539468,ai2018unbiased,zhang2022towards} using deep learning are based on Position-Based Model \cite{2015Click} (PBM) which emphasizes the key role of position as a bias factor in calculating the examination bias. According to PBM, a document has a certain probability being clicked based on the probability of it being examined and its relevance to query, where the examination depends on position and relevance depends on the features encoding the query and document. However, the examination bias is often not only dependent on the ranking position of the document in real click-through data. Therefore, recently, some works begin to consider how to add user context \cite{fang2019intervention}, clicks in the same query list \cite{chen2021adapting} and search intent \cite{SearchIntent2020} to bias factors so that the model can calculate more accurate examination bias.

In this paper, we argue that perception bias that is defined as the user's misperception of document's relevance to the query through the presentation style on SERPs, is important for figuring accurate examination bias out. Since A document has to be observed before users perceive its relevance, the examination to document can be factorized into two steps: observing and then perceiving. Obviously, the rank of document is important for it being observed by users. After the document is observed, the representation style (media type, SERP height and highlighting the hit words multiple times etc.) of it on SERPs is pivotal for users to perceive its relevance. In perception step, users often mistakenly click on irrelevant documents due to their differences in representation style.

In order to accurately calculate the perception-dependent examination bias, we first propose a Multi-Feature Integration Model (MFIM) that can integration more key bias factors that can affect user perception into examination-bias estimator. And then we mine a key factor slipoff counts that can indirectly reflects the influence of all perception-bias factors. Finally, we validate the effectiveness of MFIM on Baidu-ULTR dataset \cite{zou2022large}.

\begin{figure*}[t]
\centering
\includegraphics[width=17cm]{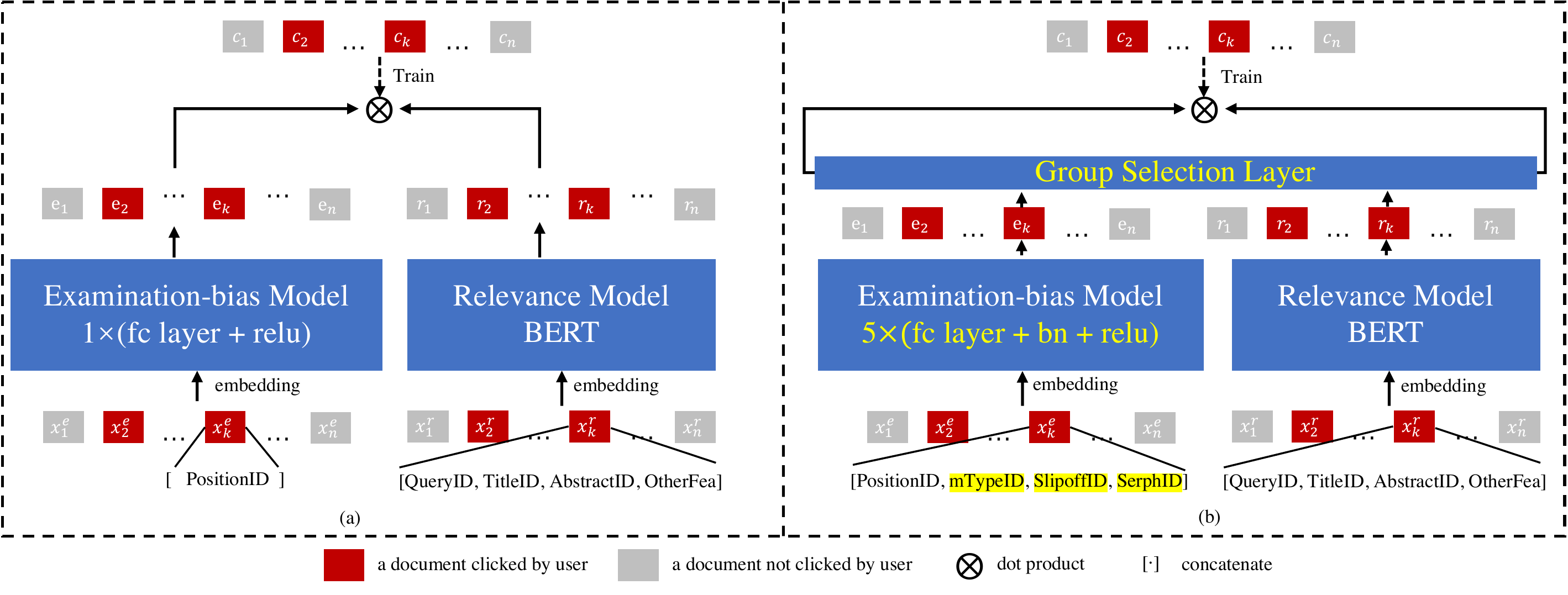}
\caption{Comparison between MFIM-based model and PBM-based models. }
\label{frame}
\end{figure*}

\section{PRELIMINARIES}
With regard to a query $q \in Q$, there is a document list ${\pi}_q$ including $n$ documents that need to be ranked according to their relevance to $q$. Let $d_k$ be a document displayed at position $k$ with the ranking features $x^r_k$ and bias factors $x^e_k$. And the probability that $d_k$ is examined by user, related to $q$ and clicked by user are denoted as $e_k \in [0,1]$, $r_k \in [0,1]$ and $\hat{c_k} \in [0,1]$ respectively. The goal of an an unbiased ranking model is to learn how to estimate accurate relevance $r_k$ from click signals $c_k \in \{0,1\}$.

According to PBM, whether $d_k$ is clicked depends on if it is examined and is related to the query, which can be formulated as:
\begin{equation}
  \hat{c_k} = {e_k} \cdot {r_k}
\end{equation}
where ${e_k}$ and ${r_k}$ can be figured out by a examination-bias model $E(x^e_k, \theta_{e})$ with parameters $\theta_{e}$ and relevance model $R(x^r_k, \theta_{r})$ with parameters $\theta_{r}$. Currently, most of ULTR methods are based on \textbf{Equation (1)} to train unbiased ranking model. Their general framework is illustrated in \textbf{Fig.1 (a)}. $E(x^e_k, \theta_{e})$ usually contains only one layer of fully connected layer (fc layer) and activation function (relu), while $R(x^r_k, \theta_{r})$ applies BERT as relevance encoder generally. When training model, the $\theta_{e}$ and $\theta_{r}$ are jointly trained by loss function
\begin{equation}
  L(c_k,\hat{c_k}) =  - \sum_{q}^{Q} \sum_{k}^{n}(c_k \cdot log \hat{c_k} + (1-c_k) \cdot log(1-\hat{c_k}))
\end{equation}
where $\hat{c_k} = sigmoid(E(x^e_k, \theta_{e}) \cdot R(x^r_k, \theta_{r}))$, while we only putting the relevance model $R(x^r_k, \theta_{r})$ to use when testing. It is worth noting that since PBM assumes that ${e_k}$ is only related to the position $k$, therefore, the $x^e_k$ in the examination-bias model only uses the position as a bias factor for calculating the 
${e_k}$ as shown in \textbf{Fig.1 (a)}.

\begin{table*}[tb]
\caption{The model performance on the expert annotation dataset with different bias factors.}
\begin{center}
\renewcommand\arraystretch{1.25}
\resizebox{\linewidth}{!}{
\begin{tabular}{c c c c c c  c  c c  }
\toprule[1.20pt]
\textbf{Method} & Position & MType & Serph & Slipoff count &\textbf{DCG@1}  & \textbf{DCG@3}  & \textbf{DCG@5}  & \textbf{DCG@10}   \\ 
\hline
MFIM(PBM-based) &$\checkmark$ & & &                                  &2.36  & 4.84  & 6.54  & 9.64 \\
MFIM &$\checkmark$ &$\checkmark$ & & $\checkmark$         &2.44  & 5.06  & 6.85  & 10.10\\ 
MFIM &$\checkmark$ & & $\checkmark$ &  $\checkmark$       &2.48  & 5.13  & 6.95  & 10.25  \\ 
\toprule[1.20pt]
\end{tabular}
}
\end{center}
\end{table*}

\section{Method}
\subsection{The Mutil-Feature Integration Model}

It takes two steps to examine a document: observing it firstly and then perceiving it. To all appearance, the PBM-based methods include the effect of the document rank on user observing document, which is not enough to figure a accurate examination bias out. For the step of evaluating document, there are many complicated bias factors except the ranking of document. For example, the media type of document significantly affect one's perception of the relevance of it to a query because different queries have different requirements for the media type of the target document.

Therefore, we argue that not only the position should be included in the bias factors for calculating the examination bias but also the other bias factors used for evaluate the one's perception bias of the relevance should. In this way, we proposed a unbiased learning to rank method named Mutil-Feature Integration Model (MFIM) that include more feasible bias factors on calculating perception-dependent examination bias. Distinctly, how to find suitable bias factors for calculating the perception-dependent examination bias is the most critical point.

\subsection{User Behaviour as Bias Factors}
One of the most naive ways to find bias factors for calculating the perception-dependent examination bias is to enumerate. We can gradually integrate the bias factors such as media type (mType) and SERP height (serph) we can come up with into $x^e_k$ and conduct ablation experiments to verify their effectiveness. However, the actions of users to perceive document relevance in the real world are too complex to enumerate all biasing factors. Therefore, we propose that the user's implicit feedback behavior after clicking the document, especially the slipoff count, can replace all factors affecting user perception of the document itself to calculate the perception-dependent examination bias. Whatever the factors for one's perception bias is, the influence of these factors will eventually be reflected in the implicit behavior of the user after clicking on the document. For example, documents misperceived by users is always have fewer slipoff count than true relevant documents. Therefore, the model can easily judge whether the user has a perception bias based on the user behavior after the click.

It is worth mentioning that although using implicit user feedback such as slipoff count does not need to use explicit document perception bias factors according to the analyses above, integrating mType, serph and slipoff count is slightly better than using slipoff count alone in practice because the explicit factors can reduce the difficulty of model training.

\subsection{Model Details}

The framework of MFIM is illustrated in \textbf{Fig.1 (b)}. There are three different points compared MFIM with the general model in \textbf{Fig.1 (a)}:

1) MFIM integrates position, mType, serph and slipoff count into $x^e_k$ while the examination bias only depends on position in general model.

2) The examination-bias model is constructed more deeply to model a more complex non-linear mapping of various bias factors affecting the perception-dependent examination bias. In addition, batch normalization (bn) is vitally important to examination-bias model since it can greatly accelerate model convergence.

3) We construct a group selection layer before calculating loss function. The role of the group selection layer is to select out a subset of $\pi_q$ randomly so that avoiding the imbalance of positive and negative samples. The subset contains one clicked document and $g-1$ document that are not clicked by users, where $g < n$. The $\hat{c_k}$ in these $g$ samples will then be fed into a softmax layer. After group selection layer, the loss function of MFIM can be formulated as
\begin{equation}
  L(c_k,\hat{c_k}) =  - \sum_{q}^{Q} \sum_{k}^{g} (c_k \cdot log \hat{c_k} + (1-c_k )\cdot log (1-\hat{c_k}))
\end{equation}
With the help of the softmax function, the training process of MFIM is between list-wise and pair-wise.

\section{Experiments}
In this section, we elaborate our experimental setting and evaluate the performance of MFIM through a real-world experiment on Baidu-ULTR dataset. 

\subsection{Experimental Set}

\subsubsection{Dataset.}
Baidu-ULTR dataset consists of two parts: 1) large scale web search sessions and 2) expert annotation dataset. The former that contains 383,429,526 queries and 1,287,710,306 documents is randomly sampled from search sessions of the Baidu search engine in April 2022. Most session contains less than 10 candidate documents with page presentation features (mType and serph etc) and user behaviors (click and slipoff count etc) of current query. The latter is also randomly sampled from the monthly collected query sessions of the Baidu search engine and the relevance of each document to the query has been judged by expert annotators who assign one of 5 labels, {bad, fair, good, excellent, perfect} to the document.

In our experimental setting, the large scale web search sessions is applied to train the ranking model and the subset of expert annotation dataset using in stage 1 is applied to validate the performance of the ranking model.

\begin{table}[tb]
\caption{Comparison with different number of fc layer in Examination-bias Model}
\begin{center}
\renewcommand\arraystretch{1.5}
\begin{tabular}{c   c }
\toprule[1.25pt]
\textbf{Method} &  \textbf{DCG@10}  \\ 
\hline
MFIM-3l & 10.05 \\
MFIM-5l & \textbf{10.16} \\
MFIM-7l & 10.14  \\ 
\hline
MFIM-5l-$g$4 & 10.16  \\
MFIM-5l-$g$6 & \textbf{10.25}  \\
MFIM-5l-$g$8 & 10.14  \\
\toprule[1.25pt]
\end{tabular}
\end{center}
\end{table}

\subsubsection{Training Details.}
The entire model is implemented by PyTorch \cite{pytorch} and trained on 8 NVIDIA A100 GPUs with batch size $16\times8$. The optimizer we used is Adam \cite{kingma2014adam} and learning rate is fixed as 5e-6. We set the maximum ranking position of candidate documents to be 10, i.e. $n=10$ and the group size $g$ is set to 6. The embedding size of every bias factor is 8. In addition, the relevance model should be pre-trained using the method whose detail can be seen at  \href{https://github.com/lixsh6/Tencent_wsdm_cup2023/tree/main/pytorch_unbias}{https://github.com/lixsh6/Tencent\_wsdm\_cup2023}.

\subsubsection{Metrics}
The Discounted Cumulative Gain (DCG) is employed to assess the performance of the ranking model. For a ranked list of $N$ documents, we use the following implementation of DCG:
\begin{equation}
  DCG@N = \sum_{k=1}^{N} \frac {G_k} {log_{2}(k+1)} \qquad
\end{equation}
where $G_k$ denotes the relevance label assigned to the document’s label at position $k$. 

\subsection{Performance of Single Model}
The performance of taking different bias factors as input to train the unbiased ranking model are shown in \textbf{Table 1}. Note, the model using position factor only at the first row can been ragarded as the model shown in \textbf{Fig.(a)}. It can be observed that when we integrate the bias factors affecting the perception bias into $x_k^r$ on the basis of the position, the ranking ability of the model will increase accordingly, which proves MFIM is outperform to PBM-based methods.

In addition, we also conduct hyperparameter experiments including how to set the number of fc layers of the examination-bias model and the group size $g$. All results can be found in Table 2

\subsection{Model Ensemble}

In order to further improve the performance of the relevance model, we used the weighted sum of the output scores of 10 models trained under different settings that we produced during the experiment as the final relevance score. The weight of each relevance model is obtained by manual search. The dcg@10 of model Ensemble on val dataset is 10.54 (10.14 on final leaderboard)

\section{CONCLUSION}
In this paper, we introduce our method on WSDM Cup 2023 Unbiased Learning for Web Search which won the $1st$ place with a DCG@10 score of 10.14 on the final leaderboard. We have the following conclusions:

1) Including the bias factors affecting perception bias except for rank position can calculate the more accurate examination bias.

2) We mine three key perception bias factors including slippoff count, mType and serph can improve the debiasing ability of the model.

\begin{acks}
This paper is supported by Tencent Machine Learning Platform Search (Tencent-MLPS). We thank everyone that offers advice to us and everyone associated with organizing and sponsoring the WSDM Cup 2023. 
\end{acks}

\bibliographystyle{ACM-Reference-Format}
\bibliography{sample-base}


\end{document}